# Thermodynamics of free bosons and fermions in the hyperball

Josep Batle[1,2] and Boris A. Malomed[3,4]

[1]Departament de Física UIB and Institut d'Aplicacions Computacionals de Codi Comunitari (IAC3), Campus UIB, E-07122 Palma de Mallorca, Balearic Islands, Spain

[2]CRISP – Centre de Recerca Independent de sa Pobla, sa Pobla, E-07420, Mallorca, Balearic Islands, Spain

[3]Department of Physical Electronics, School of Electrical Engineering, Faculty of Engineering, and Center for Light-Matter Interaction, Tel Aviv University, Tel Aviv 69978, Israel

[4]Instituto de Alta Investigación, Universidad de Tarapacá, Casilla 7D, Arica, Chile

Many-particle systems pose commonly known computational challenges in quantum theory. The obstacles arise from the difficulty in finding sets of eigenvalues and eigenvectors of the underlying Hamiltonian while enforcing fermion or boson statistics, not to mention the prohibitive increase in the computational cost with the system's size. The first obvious step in this direction is to elaborate the theory for Fermi or Bose gases without inter-particle interactions. The traditional approach to the work is with the ideal gases confined in a cubic container with impenetrable walls (in arbitrary dimensions). This approach allows one to find the particle's spectra and compute all thermodynamic quantities of the confined gas. In the present work, we consider the gas confined in a spherical container (in other words, an infinitely deep spherical potential well in $D$ dimensions), solving the corresponding Schrödinger equation using zero boundary conditions. We address the case of a finite number of particles N, either bosons or fermions, in the spherical potential box, as well as the thermodynamic limit, corresponding to $N \to \infty$. Owing to Weyl's relations, in the latter limit, the results do not depend on the shape of the box and thus approach the commonly known ones valid in the infinite space. Owing to the underlying SO($D$)

# 1. Introduction

Ideal gases of non-interacting bosons or fermions are the simplest subject of the quantum many-body theory at zero and finite temperatures, making it possible to produce analytical results that may be used as the starting point to tackle the challenging problem of systems of interacting particles [1]. In the case of spin-1/2 systems, a remarkable analytical solution is based on the free-fermion mapping provided by the Jordan-Wigner transformation [2]. It had produced the famous solution in condensed matter physics equivalent to the hydrogen atom, *viz.*, the $XY$ model of Lieb, Schultz, and Mattis [3]. The Jordan-Wigner transformation converts the Hamiltonian from an interacting spin basis into that for non-interacting fermions.

Many sophisticated methods were elaborated to ameliorate difficulties of the multi-body theory for interacting particles, such as neglecting the wave function symmetrization [4,5]. Here, we do not pursue this objective, focusing on the thermodynamics of the gas of non-interaction (free) particles trapped in the $D$-dimensional hypersphere. Recently [6], the consideration of the gas of free fermions was performed in a totally different setting, i.e., for free particles residing on lattices in an arbitrary spatial dimension, with intersite hopping amplitudes subject to a power-law decay with respect to the distance.

The study of thermodynamic quantities for particles often amounts to finding the spectrum of energy eigenvalues arising from the quantization of the corresponding modes in a $D$-dimensional rectangular box ("container"), and then taking the thermodynamic limit [7–10]. While doing that, it is usually implied that the exact shape of the container is irrelevant in this limit, according to the classical Weyl's results [11,12].

Generally, the distribution of eigenvalues of the wave equation in a bounded domain is known as the Weyl's problem. Important quantities are produced by the solution of this problem, such as the number of states with the wavenumber up to a maximum value, alongside with its derivative, the density of states. This approach is common in nuclear physics and studies of spin-polarized Fermi gases, Bose-Einstein condensates, the Casimir effect, as well as the black-body radiation. The Weyl's theorem basically tells that, in the limit of large wavenumbers, the cumulative number of states depends only on the volume of the domain, but not on its shape. Deviations from this classical result may be related to the curvature of the domain and other features. In particular, the deviations from the Weyl's theorem are relevant to the theory of quantum billiards [13].

Weyl's theory [11,12] relates three distinct notions: the number of states contained in the given spatial domain $D$, high-frequency asymptotics of the spectrum of the operator controlling its dynamics, and the geometry of domain $D$. The study of this triple relation has been a powerful catalyst for key developments in physics and mathematics alike. On the physics side, it is ultimately interwoven with the birth of the quantum theory (see e.g. [14,15]). Specifically, the Weyl's law for the Laplacian stands as an archetype for similar results in the spectral geometry, with direct implications in physics.

Starting with the $D$-dimensional Schrödinger equation with Laplacian $\Delta$, we consider large-wavenumber asymptotics of the homogenous scalar Helmholtz operator acting in hyperball $\Omega$ of radius $R$,

$$(\Delta + k_n^2)\phi_n = 0, \tag{1.1}$$

subject to the Dirichlet boundary conditions on hypersphere $\partial\Omega$ of radius $R$. In Eq. (1.1) eigenvalues $k_n$ are, as shown below, $R^{-1} j_{\nu,s}$ ($j_{\nu,s}$ is the $s$-th zero of the Bessel function of the first kind of order $\nu$). Due to the compactness of $\Omega$, eigenvalues $k_n$ form a discrete set and, denoting by $N(k_F)$ the number of eigenvalues $k_n$ below a certain wavenumber $k_F$ (the Fermi wavenumber), the asymptotic result for $k_F \to \infty$ is

$$N(k_F) \sim \text{Vol}_D(\Omega) k_F^D + o(k_F^{D-1}), \tag{1.2}$$

where $\text{Vol}_D(\Omega)$ is the volume of the $D$-dimensional hyperball with surface $\Omega$. With $k_F = \sqrt{2mE_F}/\hbar$ this is, essentially, the Weyl's law, implying $N \propto E_F^{D/2}$ or, conversely, $E_F \propto N^{2/D}$.

Several thermodynamic properties have their dependence on the dimension stemming from this result.

In the present work, we report results corresponding to finite systems confined by the hypersphere $\Omega_D$ in the $D$-dimensional space. To the best of our knowledge, such a detailed analysis was not reported previously. Other studies tackled confining potentials other than the one corresponding to the infinitely deep spherical potential box, such as the harmonic oscillator (HO), for which analytic results can be obtained readily. Also, eigenvalues for a particle trapped in a 3D ellipsoidal domain have been considered [**?**].

The HO potential is a physically important one, as it adequately models magnetic [16–18], optical [19], and magneto-optical [20] trapping potentials in experiments with ultracold atoms and Bose-Einstein condensates (BEC). Traps in the form of infinitely deep potential boxes have been created too, by means of optical techniques [21], and they are also used in many experiments with ultracold Bose and Fermi gases [22]- [29]. Actually, only 2D boxes with full isotropy are currently available (3D boxes were created by combining a hollow confining optical beam with a pair of perpendicular optical sheets [22]). In principle, spherically isotropic 3D box potentials can be built by a superposition of many hollow beams with axes isotropically converging towards the origin, resembling the illumination used for igniting the inertial nuclear fusion (with the power larger by many orders of magnitude) [30,31]. The box of dimension $D > 3$ can be created in setups which include extra *synthetic dimensions*, emulated by internal degrees of freedom (such as spin) of the particle [32,33], although, of course, it will be difficult to design a potential structure which will be isotropic in such a $D$-dimensional space with $D \geq 4$.

## 2. The Schrödinger equation for a particle inside the hyperspherical potential box

The usual treatment of the gas of non-interacting particles begins with the solution of the corresponding Schrödinger equation with the confinement potential. As said above, previous works chiefly dealt with the HO potential in $D$-dimensions. We here address the infinitely deep hyperspherical box. Thermodynamic quantities of the ideal gas are determined by the spectrum of eigenmodes in the box. The problem is far from trivial as the energy spectrum for the hyperspherical box is not available in a explicit analytical form.

Thus, we consider the Schrödinger equation,

$$-\frac{\hbar^2}{2M}\nabla_D^2 \psi + V(\mathbf{r})\psi(\mathbf{r}) = E\psi(\mathbf{r}), \tag{2.1}$$

where $M$ is the mass of the particle, and the confining potential corresponding to the infinitely deep spherical box is

$$V(\mathbf{r}) = V(r) = \begin{cases} 0 & \text{for } r < R, \\ \infty & \text{for } r \geq R. \end{cases} \tag{2.2}$$

$\nabla_D$ is the $D$-dimensional gradient operator acting on the coordinate set which can be defined in the Cartesian form, $\mathbf{r} = (x_1, x_2, ..., x_D)$, or as $D$-dimensional polar coordinates, $(r, \theta_1, \theta_2, ..., \theta_{D-1}) \equiv (r, \Omega_{D-1})$. The Cartesian and polar systems are related by expressions $r^2 = \sum_{j=1}^{D} x_j^2$, $x_j = r \cos\theta_j \prod_{k=1}^{j-1} \sin\theta_k$, $0 \leq \theta_i < \pi$, $j = 1, 2, ..., D-2$, $0 \leq \theta_{D-1} < 2\pi$. Further, the expression for the gradient operator in terms of the polar coordinates is,

$$\nabla_D = \left( \frac{\partial}{\partial r}, \frac{1}{r}\frac{\partial}{\partial \theta_1}, \frac{1}{r \sin\theta_1}\frac{\partial}{\partial \theta_2}, ..., \frac{1}{r \prod_{j=1}^{k-1} \sin\theta_j}\frac{\partial}{\partial \theta_k}, ..., \frac{1}{r \prod_{j=1}^{D-2} \sin\theta_j}\frac{\partial}{\partial \theta_{D-1}} \right). \tag{2.3}$$

and the Laplacian is

$$\nabla_D^2 = \frac{1}{r^{D-1}}\frac{\partial}{\partial r}\left( r^{D-1}\frac{\partial}{\partial r} \right) - \frac{\Lambda_D^2}{r^2}, \tag{2.4}$$

where $\Lambda_D$ is the $D$-dimensional angular-momentum operator, *viz.*,

$$\Lambda_D^2 = -\sum_{i=1}^{D-1} \frac{(\sin\theta_i)^{i+1-D}}{(\prod_{j=1}^{i-1}\sin\theta_j)^2} \frac{\partial}{\partial\theta_i}\left((\sin\theta_i)^{D-i-1}\frac{\partial}{\partial\theta_i}\right). \tag{2.5}$$

Solutions to Eq. (2.1) can be looked for in the form as

$$\psi(\mathbf{r}) = r^{-\frac{D-1}{2}}\phi(r)Y_{\ell,\{\mu\}}(\Omega), \tag{2.6}$$

where the so-called hyperspherical harmonics are

$$Y_{\ell,\{\mu\}} = N_{\ell,\{\mu\}} e^{im\theta_{D-1}} \prod_{j=1}^{D-2} C_{\mu_j-\mu_{j+1}}^{\alpha_j+\mu_j+1}(\cos\theta_j)(\sin\theta_j)^{\mu_j+1}, \tag{2.7}$$

with $C_k^j(t)$ being Gegenbauer polynomials in $t$ of degree $k$ and rank $j$, whereas

$$N_{\ell,\{\mu\}} = \frac{1}{2\pi}\prod_{j=1}^{D-2}\frac{(\alpha_j+\mu_j)(\mu_j-\mu_{j+1})!\Gamma(\alpha_j+\mu_{j+1})^2}{\pi 2^{1-2\alpha_j-2\mu_{j+1}}\Gamma(2\alpha_j+\mu_j+\mu_{j+1})}, \tag{2.8}$$

is the normalization constant. Here $(\ell,\{\mu\})=(\mu_1,\mu_2,...,\mu_{D-1})$, $\ell=\mu_1\geq\mu_2\geq...\geq\mu_{D-2}\geq|\mu_{D-1}|=|m|$, $\ell=0,1,2,...$, $m=0,\pm1,\pm2,...$ and $\alpha_j=(D-j-1)/2$ for $D\geq 3$. The so defined hyperspherical harmonics obey the angular equation,

$$\Lambda_D^2 Y_{\ell,\{\mu\}} = \ell(\ell+D-2)Y_{\ell,\{\mu\}}, \tag{2.9}$$

where $\ell(\ell+D-2)$ is the separation constant. The substitution of ansatz (2.6) in Eq. (2.1) leads to the radial equation for $\phi(r)$

$$\frac{d^2\phi(r)}{dr^2} + \left[\frac{2M}{\hbar^2}E - \frac{2M}{\hbar^2}V(r) - \frac{L(L+1)}{r^2}\right]\phi(r) = 0, \tag{2.10}$$

with $L\equiv\ell+\frac{D-3}{2}$. Inside the potential box, an obvious solution of Eq. (2.10) is the Bessel function,

$$\phi(r) \propto J_{\nu_l}\left(\sqrt{2ME/\hbar}\,r\right), \quad \nu_l = l + \frac{D-2}{2}. \tag{2.11}$$

Symmetry plays a paramount role in finding eigenstates of the Hamiltonian, a common manifestation of the symmetry being degeneracy of the energy levels. Specifically, the isotropic radial confinement makes $SO(D)$ the degeneracy group of the $D$-dimensional Schrödinger equation. As a consequence, the degeneracy for each integer value of angular momentum $l$ is given by

$$g_l = \frac{(l+D-3)!}{(D-2)!\,l!}(2l+D-2) = \frac{2l+D-2}{l+D-2}\binom{l+D-2}{l}. \tag{2.12}$$

Quantization of the energy levels of the particle is imposed by the fact that the radial wave function (2.11) must satisfy the zero boundary condition at the edge of the infinitely deep potential box, $r=R$, hence the respective energy eigenvalues are given by

$$E_{\nu_l,s} = \frac{\hbar^2}{2MR^2}\left(j_{\nu_l,s}\right)^2, \tag{2.13}$$

where $j_{\nu_l,s}$ is the $s$-th zero of the Bessel function of order $\nu_l = l + \frac{D-2}{2}$. The situation is illustrated in Fig. 1, where the well-known spherical Bessel functions for $D=3$, i.e., $\nu_l = l + \frac{1}{2}$, are displayed for $l=0,1,2,3,4,5$. Any solution of the the Schrödinger equation (2.1) for arbitrary dimension $D$ can be projected onto the 3D space, resembling an oscillating membrane.

To tackle the problem of many particles inside the $D$-dimensional ball, one needs to resort to numerical computation of the above-mentioned zeros of the Bessel function $j_{\nu_l,s}$, with $\nu_l = l + \frac{D-2}{2}$. Tables 2 and 2 show, respectively, the values of the zeros corresponding to 2D and 3D. As the order of the Bessel function for higher $D$ is either integer or half integer, the entire set of the zeros already appears in the 2D and 3D sets.

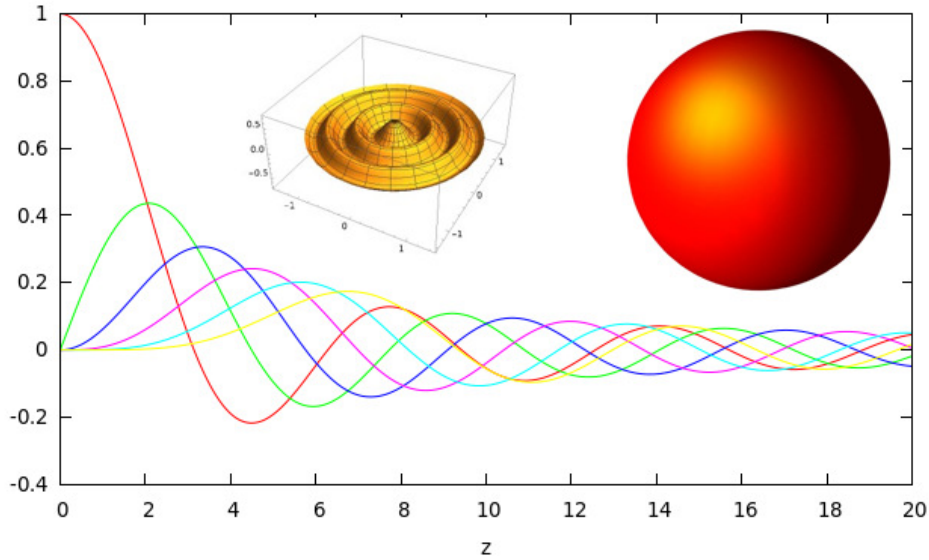

**Figure 1.** The spherical Bessel functions (for $D = 3$) as a function of $z = \sqrt{2ME}/\hbar\, r$, with orders $l \in [0, 5]$. The red $D$-ball is the space occupied by the solutions of the Schrödinger equation for a particle trapped in the infinitely deep hyper-spherical potential box. For $D = 2$, the Bessel function predicts eigenstates of the oscillating membrane or drum.

The zeros are ordered not only according to $0 < j_{\nu,s} < j_{\nu,s+1} < j_{\nu,s+2} \ldots$, but they are also *interlaced*, that is, $j_{\nu,1} < j_{\nu+1,1} < j_{\nu,2} < j_{\nu+1,2} < j_{\nu,3} \ldots$. The sorting properties of the energy eigenvalues will be of basic relevance when discussing the fermionic case.

There are approximations that work well for the upper diagonal of the spectrum, as displayed in Tables 2 and 2. There, for fixed $\nu$ and large $s$, the McMahon approximation $j_{\nu_l,s} \approx (s + \frac{1}{2}\nu_l - \frac{1}{4})\pi$ is very accurate. However, for the entries in the lower diagonal, the approximation is very poor. Some approximations are available for fixed $s$ and large $\nu_l$, but not for the entire range of zeros. Therefore, one must resort to numerical computation of the representative set of zeros, $j_{\nu_l,s}$.

When sorting the energy eigenvalues (2.13), the particular evolution of the corresponding angular momentum $l$ is apparent. As shown in Fig. 2 for the 3D and 4D cases, increasing zeros $j_{\nu_l,s}$ clearly form "bands" where the angular number $l$ varies in a rather particular way. Remarkably, in all the 2D, 3D, and 4D cases, the maximum angular momentum of the $N$-th particle is closely bounded by the curve $2.8\sqrt{N}$. Recall that $N$ stands for the total number of particles (with sorted eigenenergies). For instance, for $N = 10000$ the maximum achievable angular momentum is $\approx 2.8\sqrt{10000} = 280$.

To identify the shell structure in the energy spectrum, we can collect the number of particles between consecutive zero-values of the angular momentum, to state how many particles exists at the given shell. This is precisely what has been computed and shown in Fig. 2.

## 3. The thermodynamic study

The energy scale, determined by temperature $T$ of the gas of quantum particles trapped in the spherical box of radius $R$ is defined as $E_s = k_B T_s \equiv \hbar^2/\left(2MR^2\right)$. The gas of identical

| height $\nu \setminus s$ | 1 | 2 | 3 | 4 | 5 | 6 |
|---|---|---|---|---|---|---|
| 0 | 2.4048255576957773 | 5.5200781110286311 | 8.6537279129110112 | 11.7915344390901428 | 14.9309177084879 | 18.0710639679192 |
| 1 | 3.8317059702075112 | 7.0155866698156119 | 10.1734681350622 | 13.3236919363142 | 16.47063005087763 | 19.6158585104628 |
| 2 | 5.1356223018406383 | 8.4172441403998165 | 11.6198411721490 | 14.7959517823512 | 17.9598194949878 | 21.1169970530218 |
| 3 | 6.3801618959238984 | 9.7610231299816700 | 13.0152007216984 | 16.2234661603187 | 19.4094152264350 | 22.5827295931044 |
| 4 | 7.5883424345038040 | 11.0647094885018 | 14.3725366716175 | 17.6159660498048 | 20.8269329569623 | 24.0190195247711 |
| 5 | 8.7714838159959540 | 12.3386041974694 | 15.7001740797167 | 18.9801338751799 | 22.2177998965612 | 25.4303411542227 |

**Table 1.** Zeros $j_{\nu_l,s}$ of the Bessel functions of order $\nu = l$, with zero's number $s$. Their squared values correspond, through Eq. (2.13) to the spectrum of the two-dimensional circular potential box.

| height $\nu$ \ $s$ | 1 | 2 | 3 | 4 | 5 | 6 |
|---|---|---|---|---|---|---|
| 1/2 | 3.141592653589793 | 6.283185307179586 | 9.424777960769379 | 12.566370614359172 | 15.707963267948966 | 18.849555921538759 |
| 3/2 | 4.493409457909064 | 7.725251836937707 | 10.904121659428899 | 14.066193912831473 | 17.220755272193075 | 20.371302959287563 |
| 5/2 | 5.763459196894550 | 9.095011330476355 | 12.322940970566582 | 15.514603010886749 | 18.689036355362822 | 21.853874222270974 |
| 7/2 | 6.987932000500520 | 10.417118854737934 | 13.698023315324925 | 16.923621285213843 | 20.121806174453823 | 23.304246988939655 |
| 9/2 | 8.182561452571243 | 11.704907154570390 | 15.039664707616524 | 18.301255995954199 | 21.525417733399949 | 24.727565547835034 |
| 11/2 | 9.355812111042746 | 12.966530172774338 | 16.354709639350460 | 19.653152101821191 | 22.904550647903722 | 26.127750137225507 |

**Table 2.** Zeros $j_{\eta_l, s}$ of the Bessel functions of orders $\nu = l + \frac{1}{2}$, with the zero's number $s$. Their squared values correspond through Eq. (2.13), to the spectrum of the three-dimensional spherical potential box.

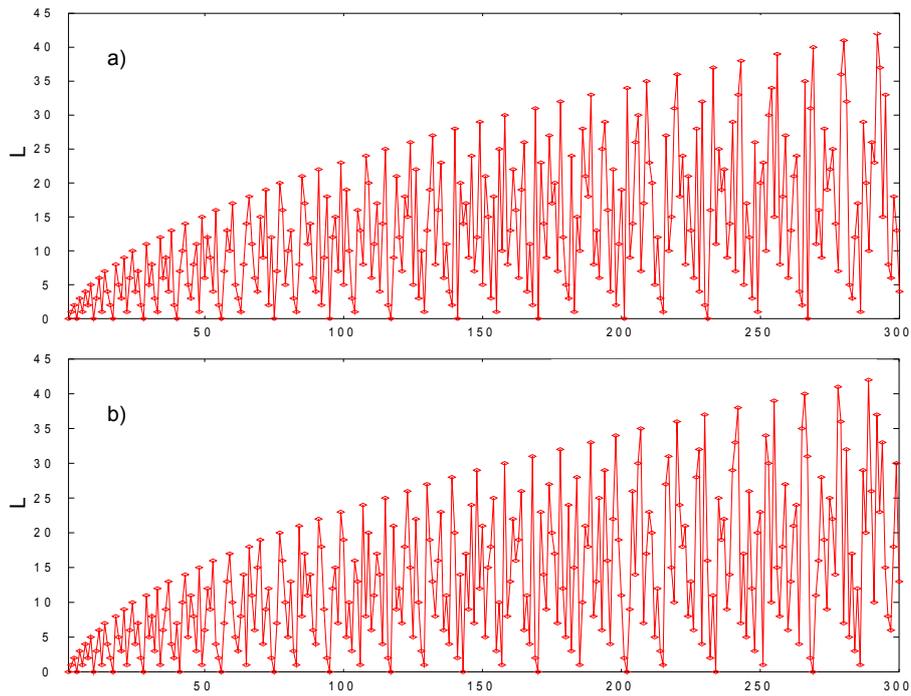

**Figure 2.** Angular momentum $L$ of the sorted values for the energy eigenvalues (2.13) for the 3D (a) and 4D (b) cases.

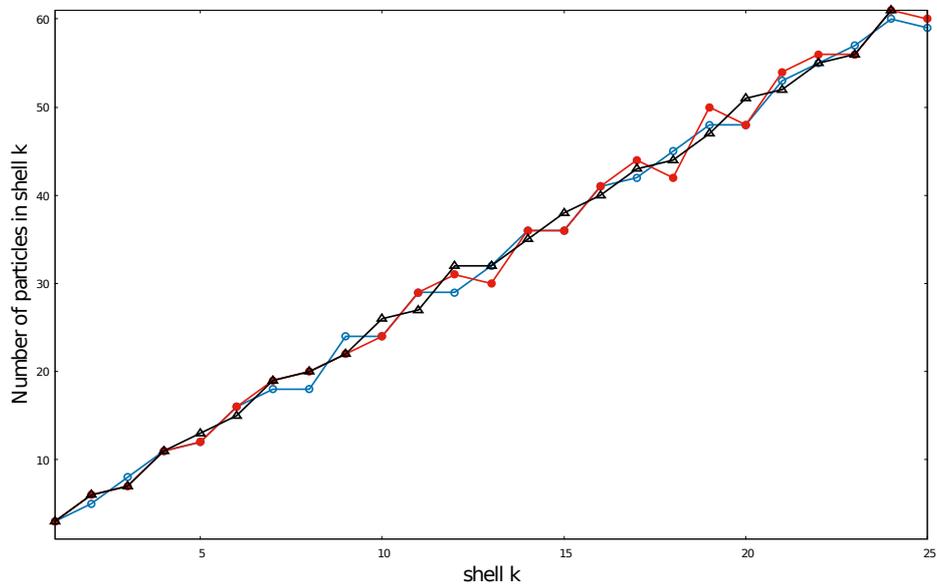

**Figure 3.** Plot of the number of states as a function of the occupied shell. Empty circles correspond to 2D, full circles to 3D and triangles to 4D. In all dimensions, the pattern is well described by a straight line, $N_{\text{shell}}(k) = 2.468 \times (\text{shell}_k)$. See text for details.

spin-polarized (bosonic or fermionic) particles is described by the many-body Hamiltonian

$$\hat{H} = \sum_{\nu_l,s} \mathcal{E}_{\nu_l,s} \hat{N}_{\nu_l,s} , \qquad (3.1)$$

where $\hat{N}_{\nu_l,s}$ is the single-mode number operator, while

$$\hat{N} = \sum_{\nu_l,s} \hat{N}_{\nu_l,s} \qquad (3.2)$$

is the total number operator.

At thermal equilibrium with temperature $T$ and chemical potential $\mu$, the grand canonical partition function is given by [10]

$$\mathcal{Z} = \text{Tr}[e^{-(\hat{H}-\mu\hat{N})/(k_B T)}] \qquad (3.3)$$

with Boltzmann constant $k_B$. The thermal average of a quantum observable $\hat{O}$ reads

$$\langle \hat{O} \rangle = \frac{1}{\mathcal{Z}} \text{Tr}[\hat{O}\, e^{-\beta(\hat{H}-\mu\hat{N})}] . \qquad (3.4)$$

In particular, relevant thermal averages are

$$N_{\nu_l,s} = \langle \hat{N}_{\nu_l,s} \rangle = \frac{1}{e^{(\mathcal{E}_{\nu_l,s}-\mu)/(k_B T)} \mp 1} . \qquad (3.5)$$

$$N = \langle \hat{N} \rangle = \sum_{\nu_l,s} \frac{1}{e^{(\mathcal{E}_{\nu_l,s}-\mu)/(k_B T)} \mp 1} \qquad (3.6)$$

$$E = \langle \hat{H} \rangle = \sum_{\nu_l,s} \frac{\mathcal{E}_{\nu_l,s}}{e^{(\mathcal{E}_{\nu_l,s}-\mu)/(k_B T)} \mp 1} \qquad (3.7)$$

where the symbol $\mp$ means $-$ for bosons and $+$ for fermions. Moreover, the partition function $\mathcal{Z}$ is related to the grand thermodynamic potential $\mathcal{W} = H - TS - \mu N$ by

$$\mathcal{W} = -k_B T \ln(\mathcal{Z}) = \mp(-1) k_B T \sum_{\nu_l,s} \ln\left(1 \mp e^{-(\mathcal{E}_{\nu_l,s}-\mu)/(k_B T)}\right), \qquad (3.8)$$

[10], and one can also determine the entropy $S$ of the gas as

$$S = \frac{E - \mu N - \mathcal{W}}{T} . \qquad (3.9)$$

In the framework of the grand canonical formulation, independent thermodynamic parameters are temperature $T$, chemical potential $\mu$, hypervolume $V$, and hypersurface $\partial S$. However, the experiments with ultracold atoms and BEC are more appropriately described by the canonical ensemble, where the total number $N$ of atoms is an independent parameter, while $\mu$ is a dependent one. For this reason, within our grand canonical formalism, it is important to calculate $N(\mu)$ at fixed $T$ for bosons and fermions and numerically invert it, obtaining $\mu = \mu(N, T)$. After that, one can investigate numerically, for bosons and fermions alike, the internal energy $E$ and entropy $S$ as functions of $T$ and $N$. One can also analyze the 2D pressure $P$ that is defined as

$$P = -\frac{\mathcal{W}}{\partial S} . \qquad (3.10)$$

It is also useful to plot the heat capacity $C_v$ for the constant volume, which is related to the internal energy $E$ by

$$C_v = \frac{\partial E}{\partial T} . \qquad (3.11)$$

In the case of bosons, one expects a peculiar behavior of $C_v$ close to the critical temperature $T_c$ of the Bose-Einstein condensation. In this connection, for the bosonic case it is interesting to produce the behavior of $T_c$ and of the BEC fraction in the boson gas. For fermions, relevant plots are $\mu(N)$ at $T = 0$ and $\mu(T)$ at fixed $N$.

## 4. Fermions

The fermionic case requires all energy eigenvalues $j^2_{\nu_l,s}$ to be sorted in the ascending order, as shown in Tables 2 and 2 for the 2D and 3D setups, respectively. The respective energy spectrum is a discontinuous "staircase" function of the form

$$N(E) = \sum_j \theta(E - E_j). \tag{4.1}$$

But there is more to it. As degeneracy $g_l$ (2.12) depends on angular momentum $l$, which is hidden in the order of the Bessel function, $\nu_l = l + \frac{D-2}{2}$, this procedure has to be performed numerically. Assuming $E_F/E_s$ to be the $n$-th sorted energy eigenvalue $j^2_{\nu_l,s}$, we count the number of contributions, taking into account the dependence of degeneracy $g_l$ (2.12) on $l$, i.e., $N = \sum_{k=1}^{n} g_l(k)$. In this fashion, we retrieve $N$ as a function of $E_F/E_s$. The ensuing results for $D = 2, 3, 4$ are presented in Fig 4

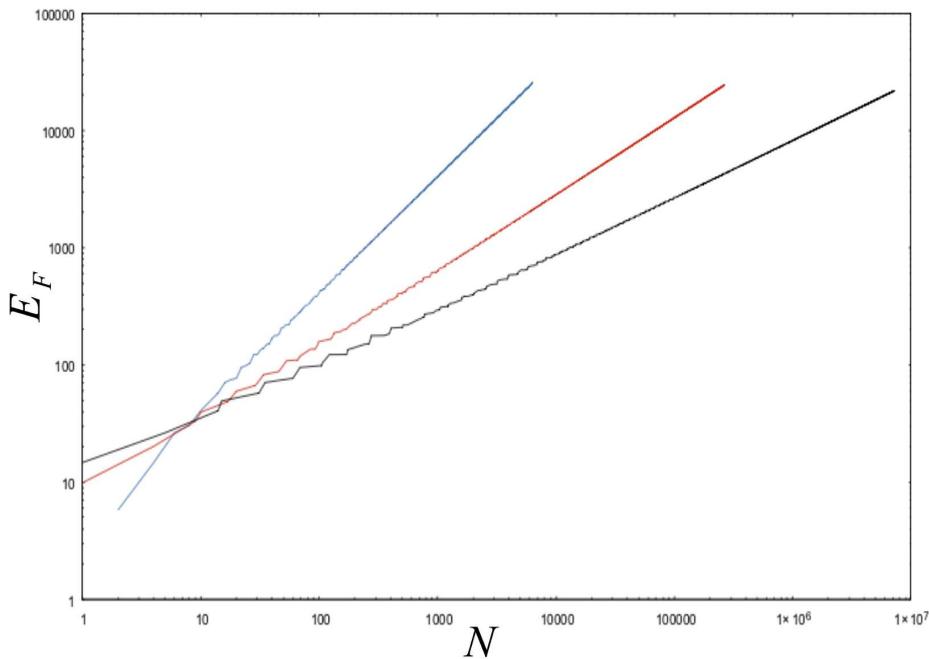

**Figure 4.** (fermions) The dependence of the Fermi energy $E_F$ (in units of $E_s$) on $N$ for the non-interacting gas of fermions trapped in the hypersphere potential box for $D = 2, 3$ and $4$ (from top to bottom). The dependence corresponds to the usual power-law $E_F \propto N^{2/D}$ for large $N$ [10], valid for the ideal Fermi gas (approximately, $E_F$ is $4N^{2/2}, 6N^{2/3}$ and $8N^{2/4}$, respectively).

The shape of the container in which the free gas is trapped is not essential, thermodynamically speaking (the Weyl's law). In the framework of the usual textbook approach, the modes corresponding to different energy eigenvalues are retrieved from the zero boundary conditions at edges of the $D$-dimensional cubic box. In the present case the use of the hypersphere trap should not lead to a different result in the thermodynamic limit. This amounts to having a smooth asymptotic expression for the Fermi energy.

The Weyl's law [11,12] secures the above statement. For the particular case of the 2D shape with area $A$ and perimeter $P$, the number of normal-mode wavenumbers in the infinitesimal interval

$(k, k + dk)$ is given by

$$dN(k) = \left(\frac{A}{2\pi}k - \frac{P}{4\pi}\right) dk, \qquad (4.2)$$

which, upon the integration, yields

$$N(k) = \frac{A}{4\pi}k^2 - \frac{P}{4\pi}k, \qquad (4.3)$$

or, in the present context,

$$N(E) = \frac{A}{4\pi}\left(\frac{2\mu}{\hbar^2}E\right) - \frac{P}{4\pi}\sqrt{\frac{2\mu}{\hbar^2}E}, \qquad (4.4)$$

hence the Weyl-like expression (4.4) is a smoothed-out approximation to the numerical results.

This type of the analysis can be extended [34] to include an additional constant (next-order) term which arises from the consideration of such geometric features as corners, curvature, and the connectivity of the 2D domain. The oscillatory behavior of $N(E)$ is the central theme of the periodic-orbit theory [35], i.e., the correction to the basic result for quantum billiards.

We have found thermodynamic quantities vs. $T/T_F(N)$, for $N = \{100, 1000, 10000, 100000\}$ and different dimensions. The dependence for a fixed dimension and varying $N$, except for $N = 100000$, falls onto the same curve for all the quantities as seen in Figs 4 (5), 4 (6), 4 (7), 4 (8), 4 (9).

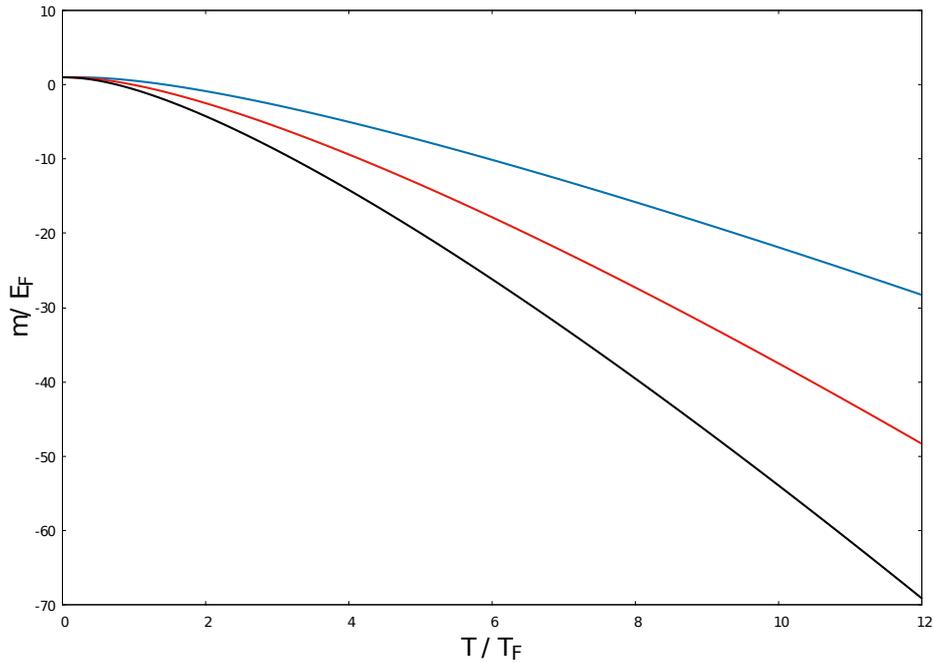

**Figure 5.** (fermions) Chemical potential $\mu$ vs. $T/T_F$ for $N = 100000$ and $D = 2, 3, 4$ (top to bottom).

The low-temperature approximation for the heat capacity goes as $C_v/(Nk_B) \approx \left(\pi^2/6\right) DT/T_F$. For $N = 100000$, the numerical slopes compared to the expected results provide upper bounds: for $D = 2$, the slope is $\approx 3.296$ vs. $\pi^2/3 = 3.289$; for $D = 3$, it is $\approx 7.47$ vs. $\pi^2/2 = 4.934$; finally, for $D = 4$ the slope is $\approx 13.7$ vs. $2\pi^2/3 = 6.579$. Large discrepancies are due to the sensitivity of the particular quantity $C_v$, and the fact that one does not truly reach the thermodynamic limit. However, the linear behavior is indeed recovered in this low-temperature regime.

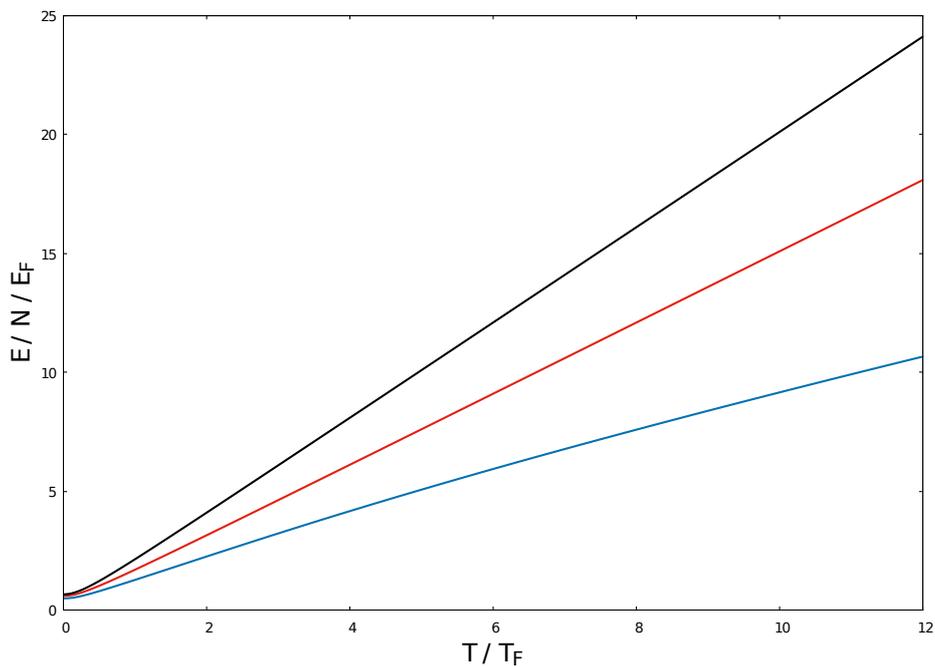

**Figure 6.** (fermions) Total energy $E$ vs. $T/T_F$ for $N = 100000$ and $D = 2, 3, 4$ (bottom to top).

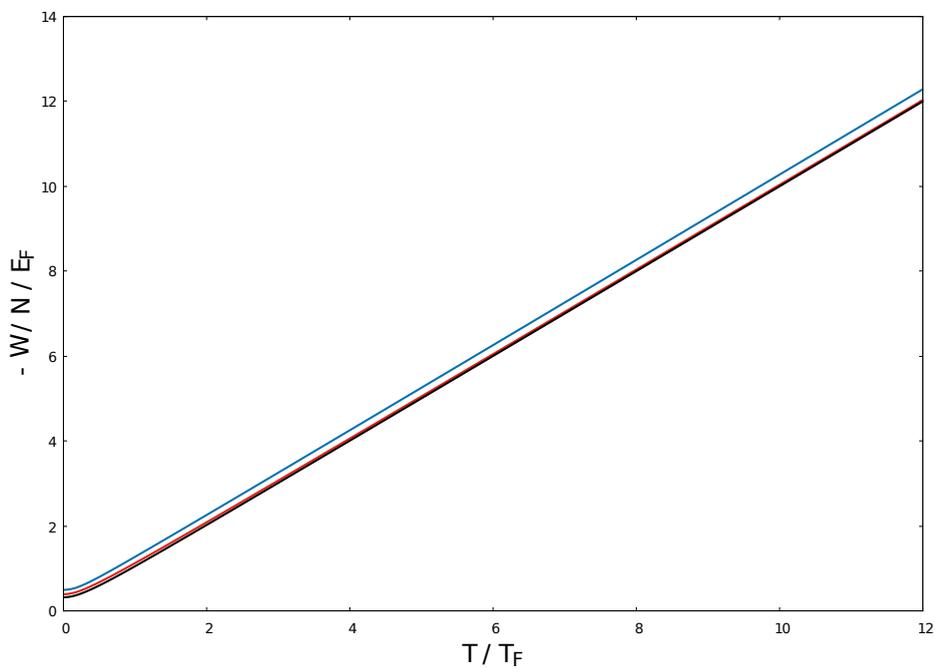

**Figure 7.** (fermions) The negative grand potential $\mathcal{W}$ (proportional to pressure $P$) vs. $T/T_F$ for $N = 100000$ and $D = 2, 3, 4$ (top to bottom).

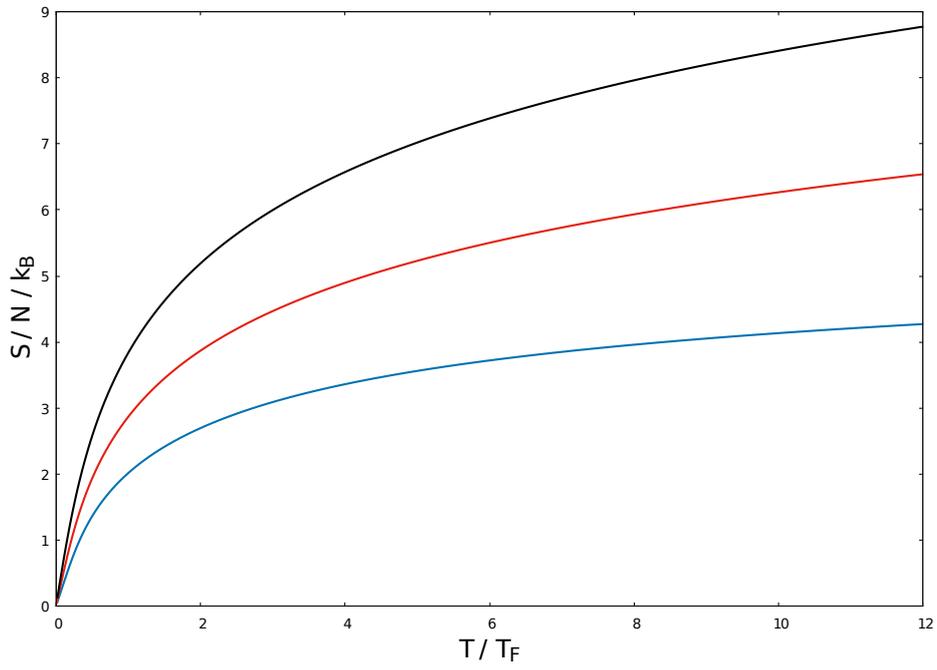

**Figure 8.** (fermions) Entropy $S$ vs. $T/T_F$ for $N = 100000$ and $D = 2, 3, 4$ (bottom to top).

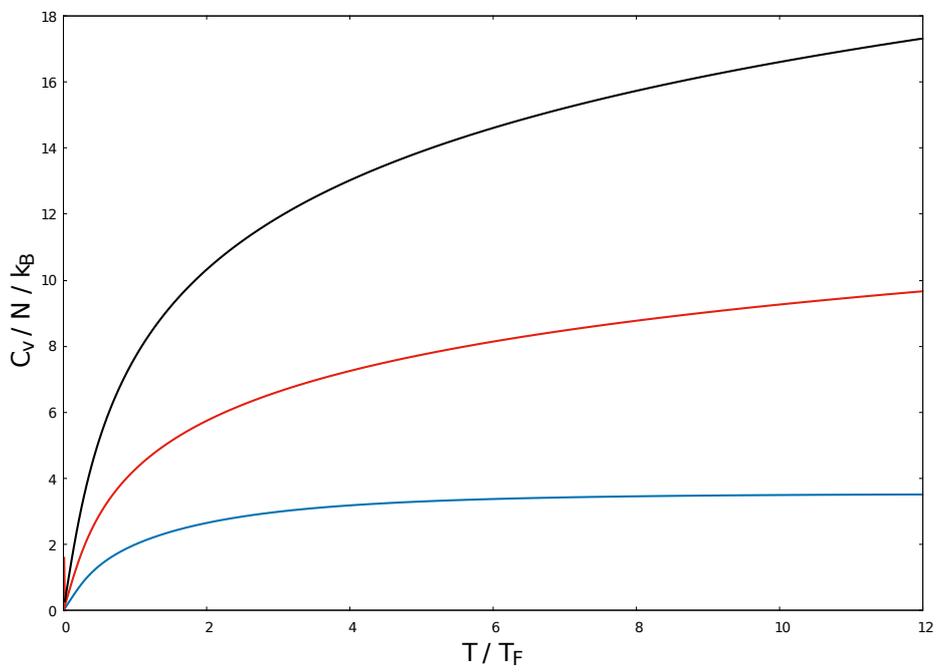

**Figure 9.** (fermions) Heat capacity $C_v$ vs $T/T_F$ for $N = 100000$ and $D = 2, 3, 4$ (bottom to top).

## 5. Bosons

We computed the dependence of the BEC critical temperature $T_c/T_s$ on the number of particles $N$, for several dimensions, numerically inverting $N = N(T_c/T_s)$. Thus, we have obtained the results displayed in Fig. 5 (10). The approximation for $N = N(T_c/T_s)$ (with $\mu = 0$), which omits $-1$ in the denominator in Eq. (3.6) and replaces the summation the by integration, retrieves an expected analytical upper bound, $T_c \propto N^{2/D}$.

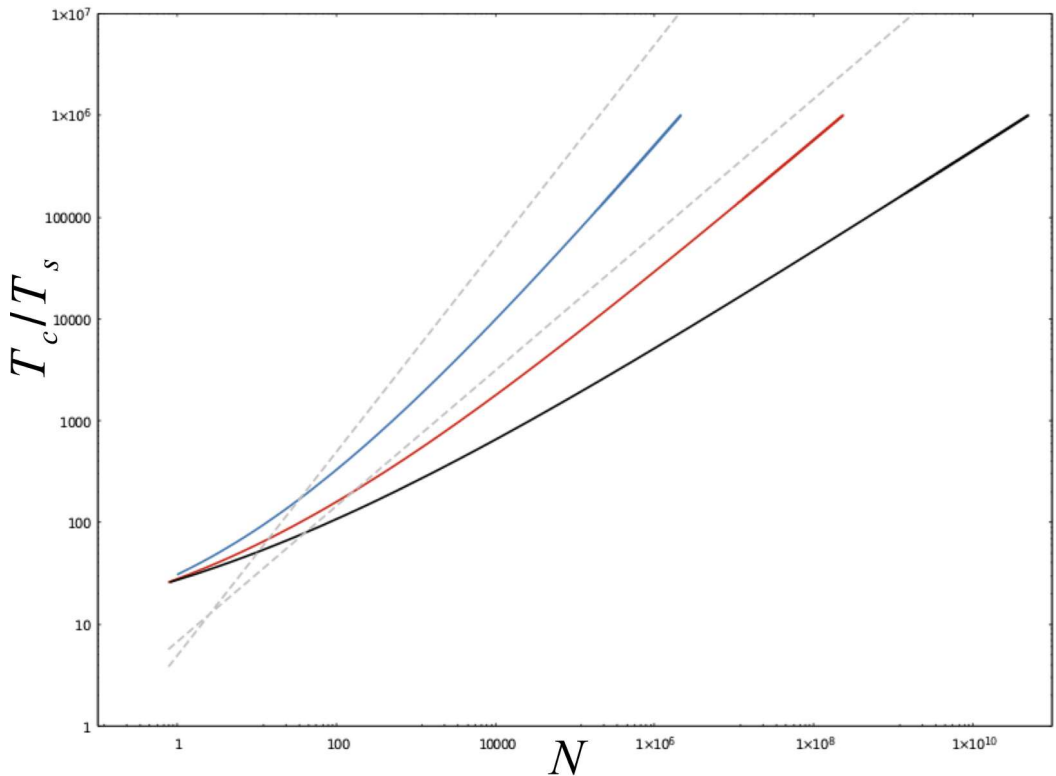

**Figure 10.** (bosons) The dependence of $T_c/T_s$ on $N$ for $D = 2, 3, 4$ (solid lines from top to bottom). Dashed lines: the analytical upper bounds for $D = 2$ and $3$, *viz.*, $\pi^2/2 N^{2/2}$ and $\pi^{5/3} N^{2/3}$, respectively. These results appear after omitting $-1$ in the denominator in Eq. (3.6), and replacing the summation by integration, with $\mu = 0$. See text for details.

Several thermodynamic quantities are found as functions of $T/T_c(N)$ for $N = \{100, 1000, 10000, 100000\}$ and different dimensions, and depicted in Fig 11 and Fig 12, as well as the corresponding BEC fractions. In this latter case, Fig. 13 displays them for 2D and 3D. As $N$ increases, the curves bend towards the origin. The last curve corresponds to the limit case $N \to \infty$, $1 - (T/T_c)^{D/2}$. The BEC fraction or any other quantity cannot numerically reach the textbook case based on the shape of the box due to the impossibility to reach the strict thermodynamic limit. For instance, $N = 2 \cdot 10^6$ pushes the curve in the 3D case only halfway between $N = 100000$ and $N \to \infty$.

We also focus on particular values of the thermodynamic quantities at $T = 0$ K and at $T = T_c(N)$ and their nontrivial dependence on dimension $D$, with emphasis on heat capacity $C_v$, thus displaying the phenomenon of the Bose-Einstein condensation for $D \geq 3$. The calculation of $C_v$ is relevant for finite boxes, as it does depend on the shape of the container (the hypersphere, in the present case). As $N$ increases, the curve for the 2D case becomes continuous and steadily approaches zero, as there is no BEC in 2D. For higher dimensions, the onset of BEC occurs.

Relevant characteristics for large $N$ and fixed $T = T_c$ are the dependences of thermodynamics quantities, such as $C_v/(Nk_B)$, on $D$. The heat capacity at the critical temperature has the dependence $\frac{D}{2}\left(\frac{D}{2}+1\right)\frac{\zeta(D/2+1)}{\zeta(D/2)}$, where $\zeta$ is the Riemann's zeta function, and the total energy $U/(Nk_BT_c)$ goes as $\frac{D}{2}\frac{\zeta(D/2+1)}{\zeta(D/2)}$ [10]. Our calculations extrapolating to $1/N \to 0$ constitute numerical upper bounds to the thermodynamic limit. For instance, in the 3D case the "spherical" heat capacity is $C_v < 2.14$, as compared to the thermodynamic-limit value $1.926$; for 4D, we have $C_v < 4.88$ and $4.384$, respectively. The 2D case is the only one for which the numerical results exactly match the expected limit values.

To additionally address the question if the results for thermodynamic quantities of the bosonic gas, obtained for finite but large $N$, are close to those corresponding to the limit of $N \to \infty$, we have performed the computations for $N = 3 \cdot 10^6$ and several values of $D$. The computed total energy and heat capacity are presented in Fig 14, as functions of $D$. In fact, our numerical results constitute an upper bound in comparison to the limit of $N \to \infty$ (lower curves), which cannot be reached in the actual numerical form. Overall, the agreement is better for lower dimensions $D$. The functional behavior, expected for $N \to \infty$ (parabolic for the total energy, and linear for the heat capacity) is well reproduced by the numerical findings obtained for $N = 3 \cdot 10^6$.

## 6. Conclusions

We have conducted the study of the ideal fermion and boson gases composed of $N$ non-interacting particles confined in a $D$-dimensional spherical potential box. Unlike the commonly known case of the cubic box, the lack of the exact analytical spectrum of energy eigenvalues in the spherical box does not make it possible to reach the true thermodynamic limit, $N \to \infty$. We, can however, pursue the study for several millions of particles, which is relevant for the ongoing experiments with Bose-Einstein condensates and Fermi gases.

The finite-size effects are apparent in basic thermodynamic quantities, such as the heat capacity. We have compared our results with the analytical ones available in the thermodynamic limit for several dimensions, finding a reasonable agreement. Our approach for the study of the thermodynamics of the non-interacting quantum gases trapped in the hyperspherical potential is, probably, the only one possible besides the textbook case of the cubic box. The study for other shapes (for instance, an ellipsoid), is out of the scope due to the underlying difficulty in computing the respective spectrum of energy eigenvalues.

In addition to recovering thermodynamic quantities expected in the limit of $N \to \infty$, we stress the fact that we dealt with the gas where particles possess a definite angular momentum. Thanks to this property, our analysis has revealed the shell structure for the spin-polarized core containing millions of particles.

Acknowledgements. The work of B.A.M. was supported, in part, by the Israel Science Foundation through grant No. 1695/22. We appreciate valuable discussions with Luca Salasnich. J. Batle acknowledges discussions with J. Rosselló, Maria del Mar Batle, Regina Batle, and Ot el Bruixot.

## References

1. Bloch I, Dalibard J, Zwerger W. 2008 Many-body physics with ultracold gases. *Rev. Mod. Phys.* **80**, 885-964. (doi.org/10.1103/RevModPhys.80.885)
2. Jordan P, Wigner E. 1928 Über das Paulische Äquivalenzverbot. *Z. Physik* **47**, 631-651. (doi.org/10.1007/BF01331938)
3. Lieb E, Schultz T, Mattis D. 1961 Two soluble models of an antiferromagnetic chain. *Ann. Phys. NY* **16**, 407-466. (doi.org/10.1016/0003-4916(61)90115-4)
4. Noack RM, Manmana SR. 2005 Diagonalization and Numerical Renormalization-Group-Based Methods for Interacting Quantum Systems, in Lectures on the Physics of Highly Correlated Electron Systems IX, AIP Conference Proceedings **789**, AIP, New York, 93-163 (doi.org/10.1063/1.2080349)

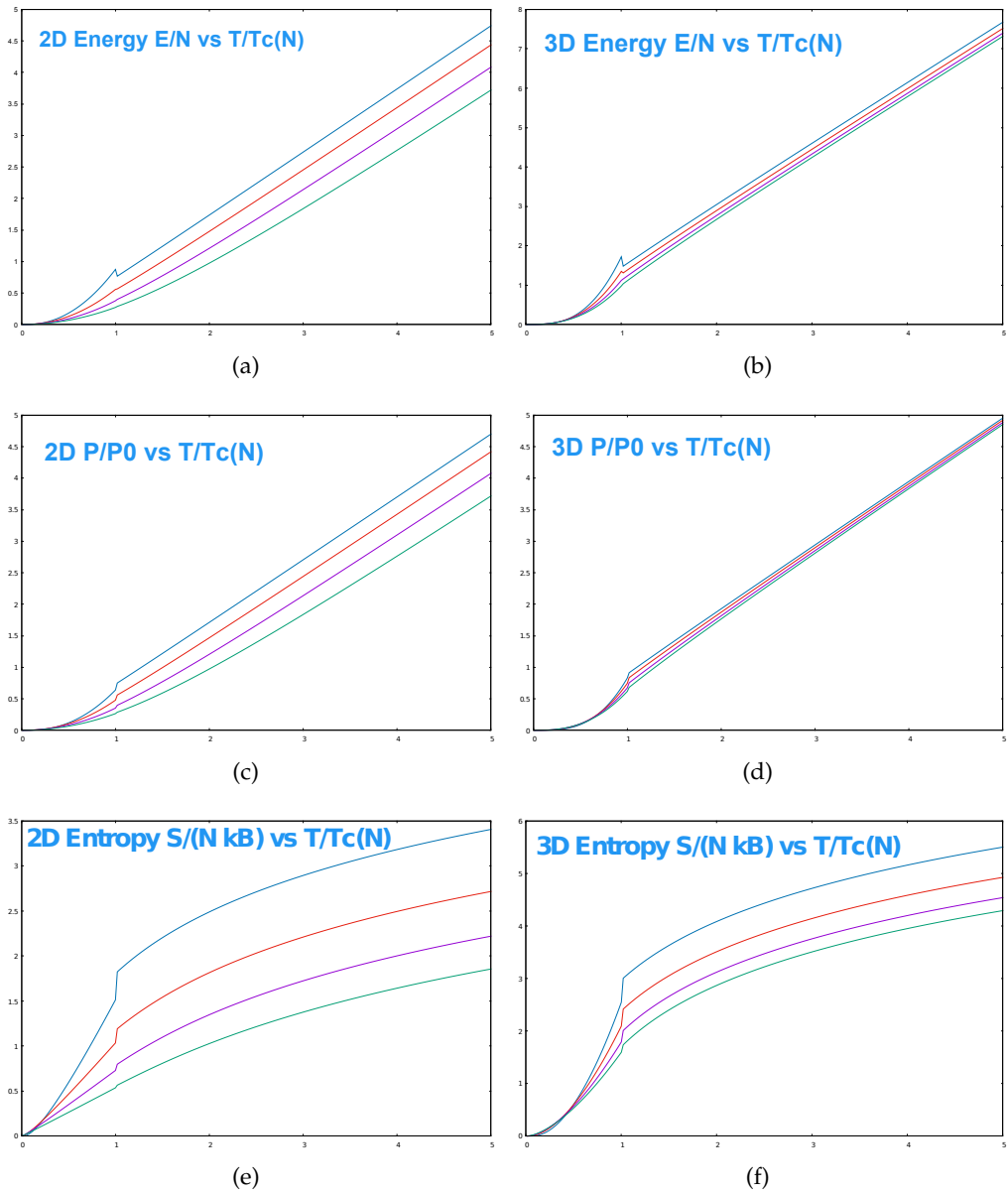

**Figure 11.** (bosons) Thermodynamic quantities vs. $T/T_c(N)$ for $D = 2, 3$, and number of particles $N = \{100, 1000, 10000, 100000\}$ (from top to bottom). (a) the 2D energy $E/N$, (b) the 3D energy $E/N$, (c) the 2D pressure $P/P_0$, (d) the 3D pressure $P/P_0$, (e) the 2D entropy $S/(Nk_B)$, (f) the 3D entropy $S/(Nk_B)$. The grand thermodynamic potential $\mathcal{W}$ is not shown, as it is the opposite of pressure $P$. The dependencies for higher $D$ tend towards merger for different values of $N$. All curves become less discontinuous as $N$ increases.

5. Schollwöck U. 2005 The density-matrix renormalization group. *Rev. Mod. Phys.* **77**, 259-315. (doi/10.1103/RevModPhys.77.259)
6. Zongping G, Guaita T, Cirac JI. 2023 Long-Range Free Fermions: Lieb-Robinson Bound, Clustering Properties, and Topological Phases. *Phys. Rev. Lett.* **130**, 070401 (doi/10.1103/PhysRevLett.130.070401)
7. Messiah A. 1961 *Quantum Mechanics*, (North-Holland, Amsterdam, Vols I and II)
8. Merzbacher E. 1970 *Quantum Mechanics*, (Wiley, New York, 2nd ed.)

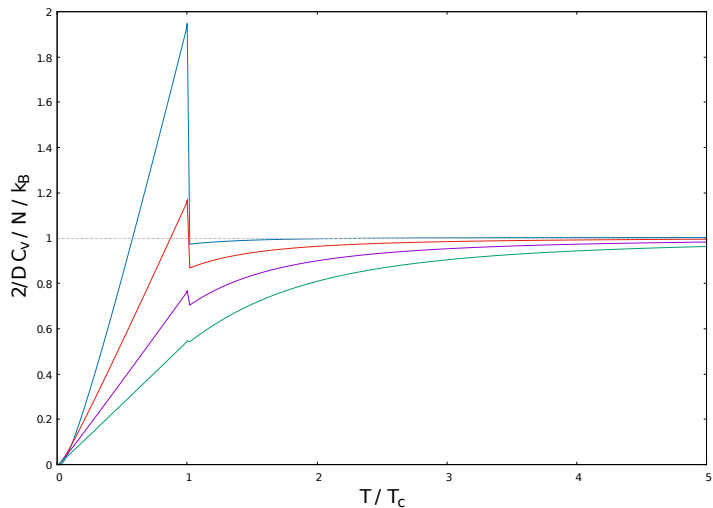

(a)

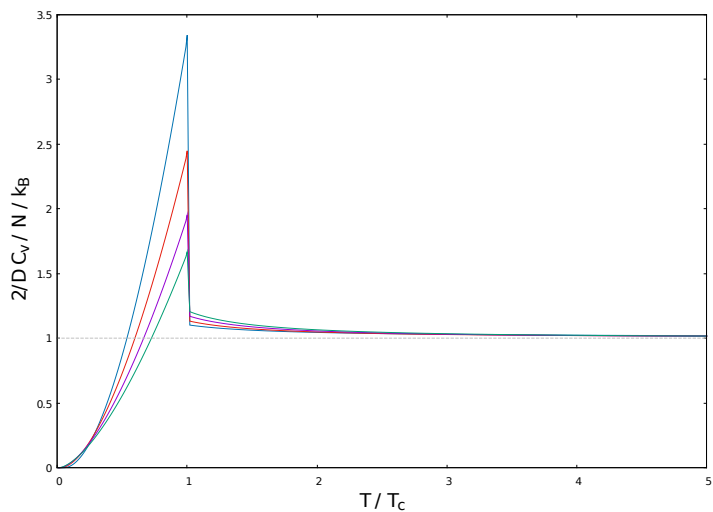

(b)

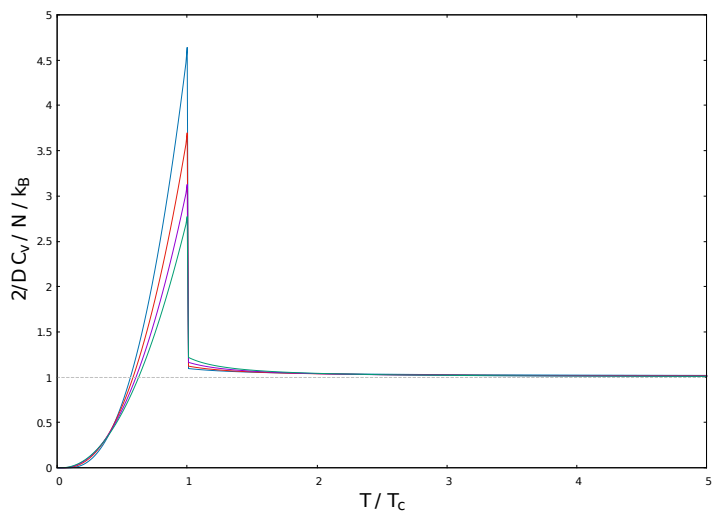

(c)

**Figure 12.** (bosons) The heat capacity $(2/D)\, C_v/(Nk_B)$ vs $T/T_c(N)$ for $D=2$ (a), 3 (b), and 4 (c), and number of particles $N=\{100, 1000, 10000, 100000\}$. As we increase $N$, the curves tend to a uniform one for $D=2$ (no phase transition), whereas for $D=3,4$, their values for $T \to T_c^{\pm}$ tend to meet each other (gestation of a BEC). For each dimension, the heat capacity decays with the increase of $N$ at $T<T_c(N)$, and it grows at $T>T_c(N)$.

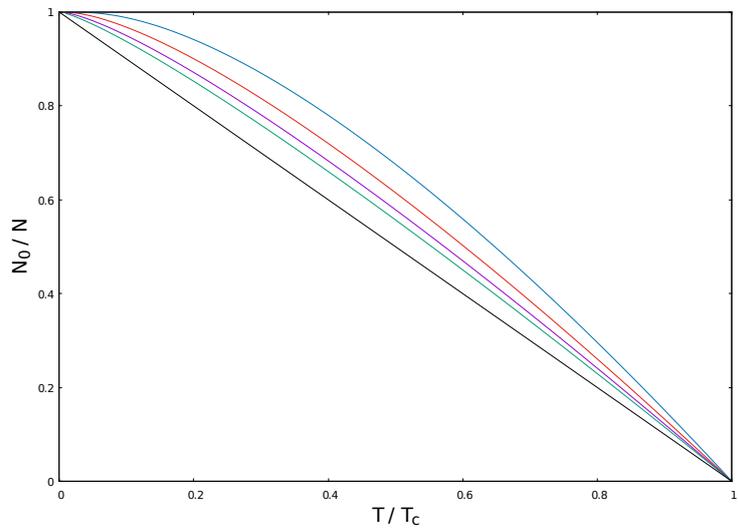

(a)

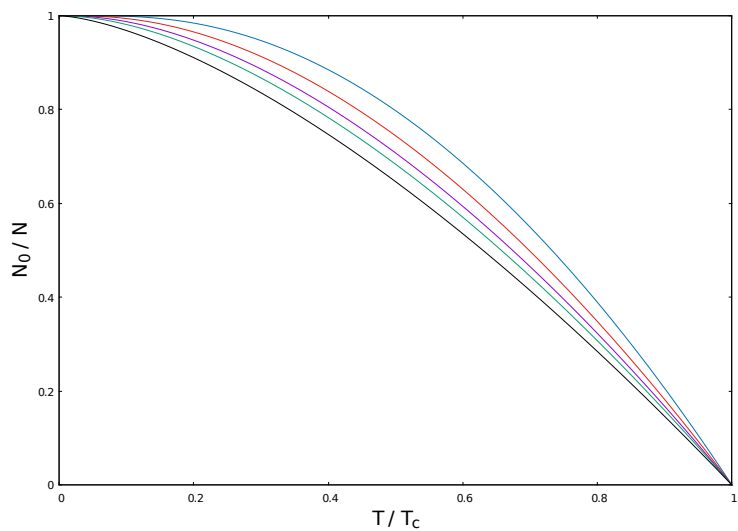

(b)

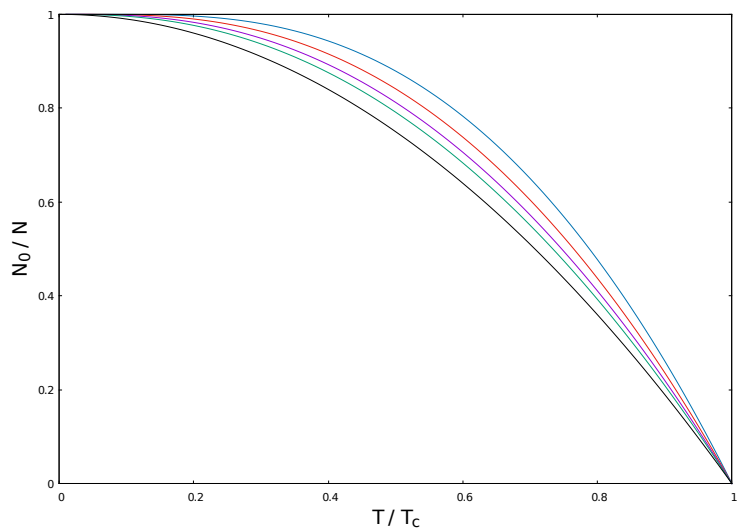

(c)

**Figure 13.** (bosons) The BEC fraction, $N/N_0$, in the bosonic gas vs. $T/T_c(N)$ for $D = 2$ (a), $3$ (b), and $4$ (c), and $N = \{100, 1000, 10000, 100000\}$. For each dimension, the curves bend towards the origin as $N$ increases. The last curve in each case is the dependence $1 - (T/T_c)^{\frac{D}{2}}$ predicted in the thermodynamic limit.

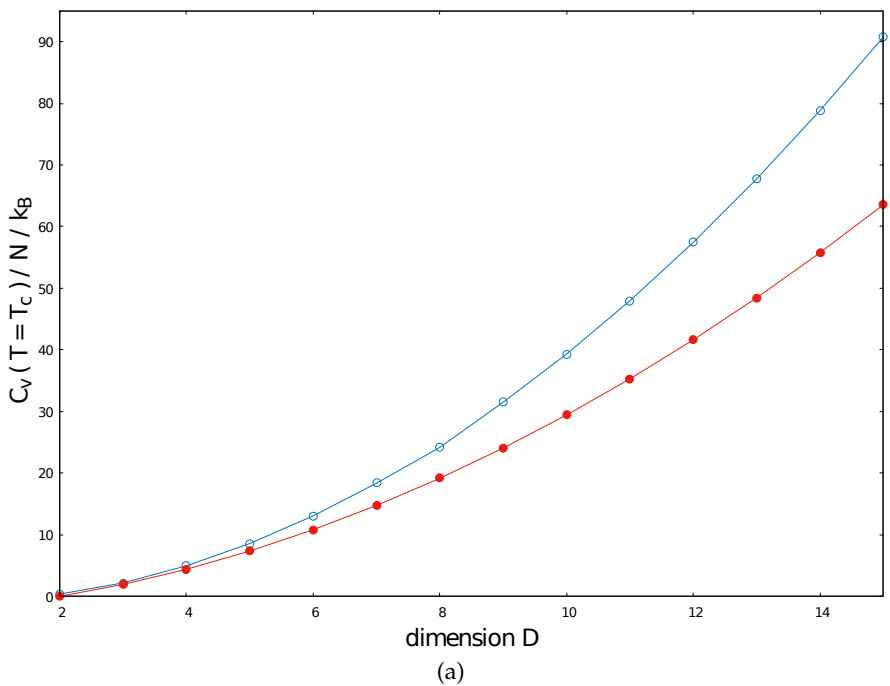

(a)

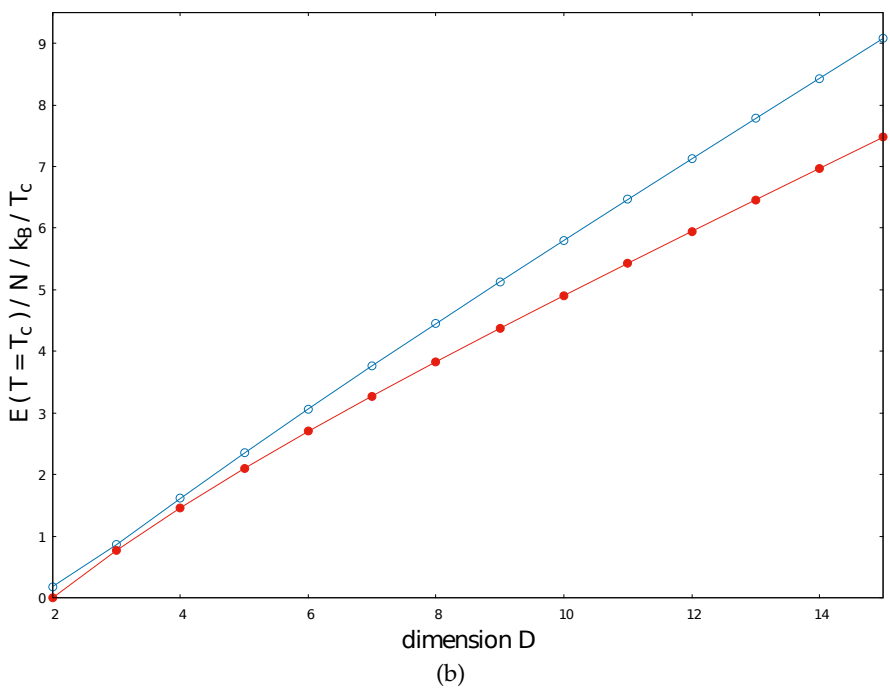

(b)

**Figure 14.** (bosons) Top curves: The heat capacity, $C_v/(Nk_B)$ (a), and total energy, $E/(Nk_BT_c(N))$ (b), of the bosonic gas at $T=T_c(N)$ vs. dimension $D$, for a fixed number of particles, $N=3\cdot 10^6$. The lower curves correspond to the ideal Bose gas for $N\to\infty$. The lines are just a guide to the eye. The results for low dimensions are in good agreement with the thermodynamic limit, see text for details.

9. Pathria RK. 1972 *Statistical Mechanics*, (Pergamon Press, Oxford)
10. Huang K. 1987 *Statistical Mechanics*, (Wiley, New York).
11. Weyl, H. 1911: Über die asymptotische verteilung der eigenwerte. *Nachrichten von der Gesellschaft der Wissenschaften zu Göttingen, Mathematisch-Physikalische Klasse* **1911**, 110-117 (https://resolver.sub.uni-goettingen.de/purl?PPN252457811-1911)
12. Weyl, H. 1912 Das asymptotische verteilungsgesetz der eigenwerte linearer partieller differentialgleichungen (mit einer anwendung auf die theorie der hohlraumstrahlung). *Mathematische Annalen* **71**, 441-479 (http://eudml.org/doc/158545)
13. Bäcker A, Quantum chaos in billiards, Institut für Theoretische Physik. (https://www.physik.tu-dresden.de/ baecker/papers/qc.pdf, last accessed: May 2024)
14. Kac, M. 1966 Can one hear the shape of a drum?. *The American Mathematical Monthly* **73**, 1-23 (doi.org/10.1080/00029890.1966.11970915)
15. Arendt W, Nittka R, Peter W, Steiner F. 2009 Weyl's law: Spectral properties of the laplacian in mathematics and physics. Mathematical analysis of evolution, information, and complexity, pp. 1-71, (Wiley, New York), (doi.org/10.1002/9783527628025.ch1)
16. Anderson, M. H., J. R. Ensher, M. R. Matthews, C. E. Wieman, and E. A. Cornell, Science 269, 198-201 (1995)
17. Davis KB, Mewes MO, Andrews MR, van Druten NJ, Durfee DS, Kurn DM, Ketterle W. 1995 Bose-Einstein condensation in a gas of sodium atoms. *Phys. Rev. Lett.* **75**, 3969-3973 (doi.org/10.1103/PhysRevLett.75.3969)
18. Bradley CC, Sackett CA, Tollett JJ, Hulet RG. 1995 Evidence of Bose-Einstein condensation in an atomic gas with attractive interactions. **Phys. Rev. Lett.** 75, 1687-1680 (doi.org/10.1103/PhysRevLett.75.1687)
19. Stamper-Kurn DM, Andrews MR, Chikkatur AP, Inouye S, Miesner HJ, Stenger J, Ketterle W. 1998 Optical Confinement of a Bose-Einstein Condensate. *Phys. Rev. Lett.* **80**, 2027-2030 (doi.org/10.1103/PhysRevLett.80.2027)
20. Raab EL, Prentiss M, Cable A, Chu S, Pritchard DE. 1987 Trapping of neutral sodium atoms with radiation pressure. *Phys. Rev. Lett.* **59**, 2631-2634 (doi.org/10.1103/PhysRevLett.59.2631)
21. Meyrath TP, Schreck F, Hanssen JL, Chuu CS, Raizen MG. 2005 Bose-Einstein condensate in a box. *Phys. Rev. A* **71**, 041604 (doi.org/10.1103/PhysRevA.71.041604)
22. Gaunt AL, Schmidutz TF, Gotlibovych I, Smith RP, Hadzibabic Z. 2013 Bose-Einstein condensation of atoms in a uniform potential. *Phys. Rev. Lett.* **110**, 200406 (doi.org/10.1103/PhysRevLett.110.200406)
23. Navon N., Gaunt AL, Smith RP, Hadzibabic Z. 2015 Critical dynamics of spontaneous symmetry breaking in a homogeneous Bose gas. *Science* **347**, 167-170 (doi.org/10.1126/science.1258676)
24. Chomaz L, Corman L, Bienaimé T, Desbuquois R, Weitenberg C, Nascimbène S, Beugnon J, Dalibard J. 2015 Emergence of coherence via transverse condensation in a uniform quasi-two-dimensional bose gas. *Nature Commun.* **6**, 6162 (doi.org/10.1038/ncomms7162).
25. Ville JL, Saint-Jalm R, Le Cerf E, Aidelsburger M, Nascimbène S, Dalibard J, Beugnon J. 2018 Sound propagation in a uniform superfluid two-dimensional bose gas. *Phys. Rev. Lett.* **121**, 145301 (doi.org/10.1103/PhysRevLett.121.145301)
26. Mukherjee B, Yan Z, Patel PB, Hadzibabic Z, Yefsah T, Struck J, Zwierlein MW. 2017 Homogeneous atomic fermi gases. *Phys. Rev. Lett.* **118**, 123401 (doi.org/10.1103/PhysRevLett.118.123401)
27. Hueck K, Luick N, Sobirey L, Siegl J, Lompe T, Moritz H. 2018 Two-dimensional homogeneous fermi gases. *Phys. Rev. Lett.* **120**, 060402 (doi.org/10.1103/PhysRevLett.120.060402)
28. Navon N, Smith RP, Hadzibabic Z. 2021 Quantum gases in optical boxes. *Nature Phys.* **17**, 1334-1341 (doi.org/10.1103/PhysRevLett.120.060402)
29. Ren T, Wang Y, Dai X, Gao X, Sun G, Zhao X, Gao K, Zheng Z, Zhang W. 2024 An efficient method to generate near-ideal hollow beams of different shapes for box potential of quantum gases, arXiv:2404.16525 (doi.org/10.48550/arXiv.2404.16525)
30. Tabak, M, Hammer J, Glinsky ME, Kruer WL, Wilks SC, Woodworth J, Campbell EM, Perry MD, Mason RJ. 1994 Ignition and high-gain with ultrapowerful lasers. *Phys. Plasmas* **1**, 1626-1634 (doi.org/10.1063/1.870664)
31. Lindl, JD, Amendt P, Berger RL, Glendinning SG, Glenzer SH, Haan SW, Kauffman RL, Landen OL, Suter LJ. 2004 The physics basis for ignition using indirect-drive targets on the National Ignition Facility. *Phys. Plasmas* **11**, 339-491 (doi.org/10.1063/1.1578638)

32. Bilitewski T, Cooper NR. 2016 Synthetic dimensions in the strong-coupling limit: Supersolids and pair superfluids. *Phys. Rev. A* **94**, 023630 (doi.org/10.1103/PhysRevA.94.023630)
33. Li Y, Zhang J, Wang Y, Du H, Wu J, Liu W, Mei F, Ma J, Xiao L, Jia S. 2022 Atom-optically synthetic gauge fields for a noninteracting Bose gas. *Light: Science & Applications* **11**, 13 (doi.org/10.1038/s41377-021-00702-7)
34. Baltes HP, Hilf ER. 1976 *Spectra of Finite Systems – Review of Weyl's Problem: The eigenvalue distribution of the wave equation for finite domains and its application to physics of small systems* (Mannheim: B. I. Wissenschaftsverlag)
35. Brack M, Bhaduri R. 1997 *Semiclassical Physics* (Reading MA: Addison-Wesley)